# Pipeline-Centric Provenance Model

Paul Groth, Ewa Deelman, Gideon Juve, Gaurang Mehta
USC Information Sciences Institute
4676 Admiralty Way, suite 1001
Marina del Rey, CA 90292
+1-310-822-1511
{pgroth, deelman, gideon, gmehta}@isi.edu

Bruce Berriman
Infrared Processing and Analysis Ctr.
100-22 Caltech
Pasadena, CA 91125
+1-626-395-1817
gbb@ipac.caltech.edu

## ABSTRACT
In this paper we propose a new provenance model which is tailored to a class of workflow-based applications. We motivate the approach with use cases from the astronomy community. We generalize the class of applications the approach is relevant to and propose a pipeline-centric provenance model. Finally, we evaluate the benefits in terms of storage needed by the approach when applied to an astronomy application.

## 1. INTRODUCTION
Provenance is commonly defined as the origin, source or history of the derivation of some object. For scientists, provenance of scientific results would indicate how results were derived, what parameters influenced the derivation, what datasets were used as input to the experiment, etc. In other words, provenance of scientific results would help reproducibility [1, 2]--a fundamental tenet of the scientific method.

Data provenance has recently attracted significant interest in several areas, including e-Science and grid computing, databases, visualization, digital libraries, web technologies, and operating systems [3] [4]. Of particular interest is the provenance of data generated by scientific workflows [5]. Today, as data are generated automatically through the execution of complex, inter-related processes, it is ever more difficult to interpret the results.

While data are being processed, provenance information can be automatically captured and then stored in a provenance store. The resulting derived data products (both intermediate and final) can also be stored in an archive, with metadata about them stored in a metadata catalog and location information stored in a replica catalog. Thus, in the context of computer systems, the provenance of a data product is the process that led to that product, where process encompasses all the derivations, datasets, parameters, software and hardware components, computational processes, digital or non-digital artifacts that were involved in deriving and influencing the data product.

In theory such provenance could be extremely large, however, in practice, detailed provenance information is not required by end users, since their needs tend to be limited to specific tasks, such as experiment reproducibility or the validation of an analysis.

In this paper we focus on the provenance of data derived by scientific/computational workflows [5, 6]. Computational workflows have become a useful tool in conducting complex scientific analyses. They provide a framework that can be used to compose data processing and simulation codes developed by different scientists. At the same time, workflows have also become a useful representation for managing the execution of large-scale computations. The workflow representation not only facilitates overall creation and management of the computation but also builds a foundation upon which results can be validated and shared. Computational workflows are used to analyze data from instruments, to simulate complex phenomena, to mine large geographically distributed data sets, and perform other sophisticated computations.

In this paper we argue that in the case of data derived through scientific workflows, we can minimize the amount of provenance that needs to be stored in order to provide information about how data was derived and to enable reproducibility.

The contributions of this paper are:

1. A simplified approach to provenance capture for a class of deterministic applications.
2. A definition of application characteristics where this approach is applicable.
3. Preliminary results showing storage usage improvement.

The rest of the paper is organized as follows. The next section illustrates the need for provenance in astronomy applications and describes the Montage application [7, 8] [9] that motivated our work. Section 3 generalizes the class of applications that can benefit from our proposed approach. Section 4 describes our pipeline-centric provenance model. Initial evaluation of the approach in the context of the astronomy application Montage is shown in Section 5. Related work is discussed in Section 6, followed by conclusions in Section 7.

## 2. Provenance in Astronomy Applications
In this section we describe the use cases that motivated our work.

Cases 1 through 3 are derived from surveys of users of the data archives at the California Institute of Technology's Infrared Processing and Analysis Center (IPAC). The fourth use case derives from the expected rapid observing cadence of the Large Synoptic Survey Telescope (LSST). The use cases demonstrate the need for both metadata (data descriptions that assign meaning to the data), and data provenance (information about how data was derived). Both metadata and provenance are critical to the ability to interpret a particular data item and thus vital to the scientific process as it is conducted in-silico.

### 2.1 Use Case 1:
A researcher is studying the dynamics of a cluster of galaxies and wants to know what imaging data are available in archives and in the literature. The researcher wants to know the following about available images:

1) What images are available for the cluster of interest that contain wavelengths between 0.3 and 2.2 microns, and have a spatial extent of 3 x 3 degrees.
2) For those images obtained by surveys, what are the sensitivities of the surveys at these wavelengths?
3) How are the images constructed:
   a) Are they data sets released by missions, or by individual astronomers?
   b) If data are released by individual astronomers: What are the telescope, bandpass, instrument, and time?
   c) What processing package and version/algorithm(s) were used to process the images? Where can the packages be found?
   d) Are the raw data and calibrators available? Where are they located?
   e) Where in the cluster of interest are all these images?
   f) What are the image footprints on the sky? What are the pixel scales? etc.

Clearly not all questions relate to the provenance records of the images that the scientists searches. As in most cases, the questions posed relate to both the metadata about the original/raw data and the metadata that was associated with the processed images. One could argue that questions 1, 2, 3 a, b, d, e, and f are metadata queries establishing the existence of data products and their basic attributes. However, questions 3 c and d are provenance questions, which aim to establish how the images where constructed during processing and to support reproducibility as in the case of question 3d. Often, such information is not released along with the image data but is necessary for the interpretation of the objects seen (or not seen) in the cluster images.

## 2.2 Use Case 2

A researcher wants to use multi-wavelength images of the Taurus Dark Cloud to construct a catalog of very faint protostar candidates (at the plate limit) to support an observing proposal. Knowing the details of the image processing is absolutely crucial for a meaningful analysis.

The researcher needs to know:

1. All the image data available online between 0.5 microns and 25 microns.
2. If these are "primary" mission products:
   a. What processing algorithms were used?
   b. How was the original data calibrated?
   c. Are the original files and algorithms available?
3. If the images are created from other products:
   a. How have the original products been processed?
   b. Has the data been averaged and reprocessed in space and time?
   c. Have the input image parameters (projection, sampling, orientation) been changed to make these products?
   d. How well have they been calibrated?
   e. How were calibration offsets between images handled?
   f. Are the algorithms available?
   g. For ground-based data, how have backgrounds been rectified or removed?
4. Where are the original images?
5. What are their limiting sensitivities?
6. Have artifacts in the images been identified?

In this scenario, most of the questions relate to the provenance of the data. Only questions 1, 4, 5, and 6 relate to image metadata.

## 2.3 Use Case 3

A researcher wants to carry out a spectroscopic study of the abundances of quasars using only available echelle data in the archives or the literature. Some of the questions that need to be answered are:

1. What echelle spectra are accessible from 0.3 microns to 2.2 microns in ground-based archives.
2. If the spectra are reduced, or if they are in the literature:
   a. What code/algorithm was used to reduce them?
   b. Are atmospheric absorption effects removed?
   c. Has any flux calibration been attempted?
   d. Are trace profiles available for each of the orders? Point Spread Functions? Signal-to-noise? Line profiles?
3. If the spectra are not reduced:
   a. What is the calibration data and where is it dark, biased, traced, focused, or flat?
   b. What calibration lamps were measured?
   c. Is the calibration data available?
4. For all data, what is the:
   a. target, position, exposure start/stop time, wavelength range, exposure time, program info (PI, etc), telescope, instrument, grating, weather logs on date of observation.

In this scenario, questions 1 and 4 refer to metadata and questions 2 and 3 refer to provenance.

## 2.4 Use Case 4

The LSST is expected to begin operations in 2015 [10]. About 90% of the observing time will be devoted to a deep-wide-fast survey mode that will observe a 20,000 square degree region about 1,000 times. The rapid cadence of this program will produce about 30 TB of data per night, leading to a total of 60 PB of raw data, and 30 PB of metadata over ten years of operations. The total data volume after processing will be several hundred petabytes. Permanent archiving of this volume of data is not feasible, so the efficient recording of provenance is a crucial part of LSST's data management plan.

## 2.5 Montage—A commonly used astronomy application

Montage, developed at Caltech, is an application that constructs custom science-grade astronomical image mosaics on demand based on several existing images. The inputs to the workflow include a "template header file" that specifies the mosaic to be constructed, and several input images in standard FITS format (a file format used throughout the astronomy community) [11]. Input images are taken from archives such as 2MASS [12]. The input images are first re-projected to the coordinate space of the output mosaic. The re-projected images are then background rectified and co-added to create the final output mosaic. Figure 1 shows the structure of a small Montage workflow using vertices to represent tasks and edges to represent data dependencies between tasks. Montage workflows typically contain a large number of tasks that process large amounts of data. For example, a workflow to generate the 2 degree square mosaic of 2MASS images centered around the celestial object M17 would contain approximately 1,000 individual tasks.

## 3. CLASS OF APPLICATIONS

Montage is an example of a class of well-specified deterministic applications that are common in science. These applications usually consist of a series of codes (i.e. components) connected together to perform large-scale analysis routines. Other examples of this class of application include: seismic hazard analysis for earthquake forecasts, analysis of large-scale social networks, analysis of the epigenomic properties of DNA sequences, searching for gravitational waves in interferometer data, and many others. These applications have a number of characteristics that can be taken advantage of to enable the reproducibility of results and the determination of provenance. These characteristics are as follows:

1. The application is deterministic. Repeating the application with the same inputs produces the same outputs.
2. The application is automated. The application does not require human intervention to execute.
3. The application is not monolithic i.e. the application is broken up into many different components that are connected together in a workflow.
4. The application is self-contained. By this we mean, that the application and all its components can be easily assembled in one location. For example, the components of Montage can be assembled in one directory.
5. The application does not require any specialized hardware to function.
6. The application uses data from well-known, well-documented sources. In astronomy, for example, significant effort is deployed in documenting the functionality of the telescopes and satellites that provide source data.
7. Source data is well preserved, archived systematically, and can be readily accessed. For example, the Sloan Digital Sky Survey provides direct access to archived image data at http://das.sdss.org.

The last characteristic is optional in our approach. However, if applications rely on such sources, our approach can optimize the data provenance storage further (Section 5).

Not all applications have these characteristics. For example, some applications rely on services provided by third parties and thus the components of the application cannot be assembled in one place. Other applications require direct interaction with a human. For example, a human's intervention might be necessary to steer a computational simulation. Still others might not be deterministic. For example, the application could be dependent on a true random number generator i.e. one initialized by a physical process.

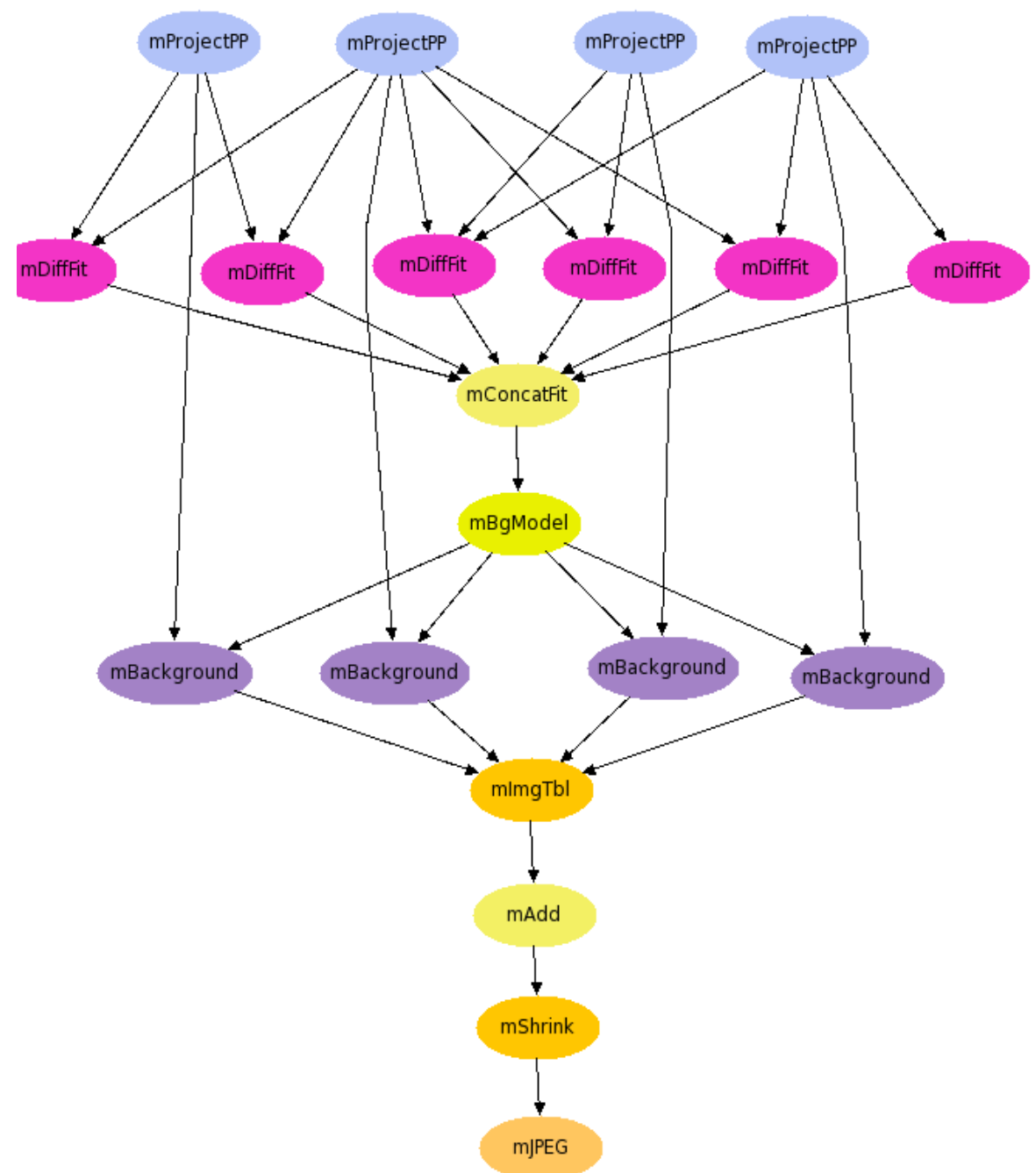

Figure 1: Small Montage Workflow.

However, while these characteristics are not universally applicable, they do describe a wide variety of important scientific applications as noted above. We now present a model for provenance that takes advantage of these characteristics.

## 4. PIPELINE-CENTRIC PROVENANCE MODEL

To determine the provenance of an application's output, one needs to be able to ascertain the relationship between the steps involved in generating the output, how those steps executed, and what data each step used during execution. This information can be modeled as a graph (Figure 2) that links the output data to the process/component that generated it, which in turn is linked to its input data, which is likewise linked to another component and so on. Thus the goal is to obtain such a *provenance graph* that accurately reflects the execution of the application in question. It is important to note that users may ask provenance questions about any portion of the graph not just the output. For example, they may ask for the provenance of a particular intermediate data product.

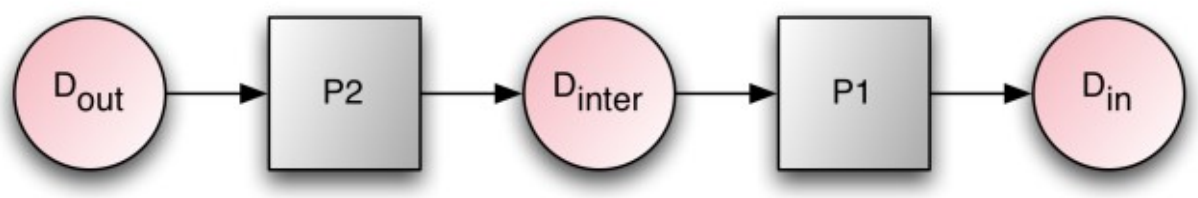

**Figure 2: A basic provenance graph.**

One approach to obtain this graph is to instrument the application to capture all steps and all the data resulting from those steps, including intermediate data. However, for scientific applications, storing intermediate data is not practical because of storage constraints. Another downside to this approach is the need to instrument the application in order to track data flow, which for many applications is infeasible due to the usage of legacy codes. However, because the applications we consider have the 7 characteristics listed in Section 3, we can take a new approach that circumvents these issues.

The approach we adopt is to leverage the workflow or pipeline used to *define* the application as the core of our model. The pipeline defines the nodes and edges in the provenance graph under the assumption that the pipeline defines all inputs and outputs of every component. (Later we discuss how to deal with conditional branches.) This inversion is possible because the application is deterministic (Characteristic #1). The pipeline itself is not sufficient to answer provenance queries, in particular, about intermediate data. For example, which data products led to $D_{inter}$ being as it is (in Figure 2).

To answer these queries, we need to be able to re-execute the pipeline to duplicate the original run. This requires the following information in addition to the pipeline:

- The original input data.
- The executables corresponding to each component defined in the pipeline.
- The parameter settings for each component.
- The execution environment for running the application.

With this information, we can reproduce any intermediate data product. Thus, intermediate data in our approach is treated as virtual data [13, 14]. Then the procedure to determine the provenance of any $D_{inter}$ would be to determine the subgraph of the workflow that is responsible for $D_{inter}$, and re-execute that subset. During the re-execution phase, one could also store all intermediate data products and return those as part of the answer to the provenance query. For workflow management systems that support conditions as part of their workflow language, this re-execution approach could be used to determine when a particular conditional branch was taken by re-executing up to that condition. Obviously, if execution overhead is of concern, intermediate data products can still be stored. An interesting test of this approach would be to pick the set of intermediate data products that would optimize re-execution for determining provenance. In other words, which data are cheaper to store than to regenerate.

One important question is whether this approach can accurately deal with determining the provenance of errors. In workflow systems such as Pegasus [15], errors in data are explicitly modeled as outputs (including stderr files). Thus, we can trace back through the workflow to determine which component is responsible for the error. Furthermore, because our approach specifically captures the execution environment, for almost all non-hardware related errors, we can determine the exact situation in which the error occurred.

Thus far we have discussed our general pipeline-centric model and its requirements. We now present the realization of this model.

## 4.1 A Pipeline-Centric Provenance Package

Our model is realized as a directory containing a workflow, a set of files in subdirectories, and a manifest that ties the contents together. The directory can be compressed as a zip or tar file to create a package describing the provenance of the experiment. This approach to packaging is common. For example, both Open Office and Microsoft Office use it for storing office documents (see the Open Document Format and Open Packaging Convention respectively). Additionally, myExperiment Packs [16] and Kepler KAR [17] files use a similar technique.

In our approach provenance packages are WHIP bundles (http://www.whipplugin.org/). The manifest of a WHIP bundle is an XML file conforming to the Atom Feed Schema (http://atompub.org/rfc4287.html). Atom is a widely used format for syndicating content over the web. An Atom feed consists of a a series of entries, each of which contains a list of categories. A WHIP manifest file contains a single entry. The categories in a WHIP manifest file point to the various contents of the bundle. Importantly, categories can point to both objects within the bundle and remote objects. The manifest for a provenance package includes:

- Metadata such as the creator of the bundle, the date of creation, and the workflow format
- The workflow description
- Input data
- Output data
- Virtual machine characteristics and the VM image location

The VM image contains software needed to execute the workflow-based application. In our example this includes: Globus [18], Condor [19], Pegasus and application binaries.

We expect that the virtual machines are configured as they were used in the execution of the workflow. Thus, they should contain all the necessary libraries for running the codes required by the workflow. The use of virtual machines is fundamental to our approach as it allows the entire execution environment to be captured, thus allowing for exact replication.

Figure 3 shows a portion of the contents of a WHIP manifest. The category "entrypoint" refers to the file containing the workflow described using the Pegasus [20] DAX format [21]. Note that the DAX will also contain the parameter settings for the workflow.

The category "VM" contains a URL to the VM image that was used in the execution of the workflow. The VM images can be quite large as can be seen in Table 1, but they can be reused by a number of (in this case) Montage workflows. Thus it may be beneficial to include only references to them. This reliance on virtual machines is enabled by the notion that the application does not run on specialized hardware (Characteristic 5). Another approach would be to store the application codes in the WHIP rather than in the VM image. Storing application codes in the WHIP would potentially provide efficient re-execution in cases where it is not necessary to load the VM; for example if the current execution environment is suitable for running the workflow.

The category “inputfile” refers to all the inputs required to rerun the workflow. When using input data that is stored long-term in an archive the actual input files can be omitted from the bundle and instead URLs or a metadata query to this data can be provided.

The final category “outputfile” refers to the outputs of the execution of the workflow. If the outputs of the workflow are large, they can be omitted from the bundle and the workflow can be re-executed to reproduce them.

For applications that follow the characteristics described in Section 3, a WHIP bundle containing all the information described above, provides all the necessary information to re-execute the experiment, and determine the provenance of the data. However, while such a package is comprehensive, it also requires significant storage space. Example bundles are located at: http://pegasus.isi.edu/workflows/montage/.

## 4.2 Storage Efficiencies through References

By taking advantage of the characteristics of the applications we consider, and by using the functionality of the WHIP bundle to refer to external locations, the size of a provenance package can be significantly reduced, as demonstrated in Section 5. In the extreme case, we imagine that the entire bundle would only contain metadata, a workflow description, and references to input data and VM images. This is under the assumption that all input data, all virtual machines, and all codes are stored in a remotely accessible archival repository.

While it is not the case at the moment, many scientific fields are beginning to store data in curated archives. As previously mentioned, sky survey data is available from such a repository. In addition, the scientific community that studies climate change has set up a network of data centers for topics ranging from biodiversity to glaciology (http://www.ngdc.noaa.gov/wdc/list.shtml). Besides data sets, there are a number of national software repositories for scientific computing codes [22]. Finally, Amazon provides a number of preconfigured virtual machine images for use on their cloud. These preconfigured virtual machines are a step towards an accessible archived library of virtual machines.

In the next section, we show how the pipeline provenance model can reduce the amount of storage needed for provenance through the use of re-execution and references.

```
<?xml version="1.0"?>
<entry xmlns="http://www.w3.org/2005/Atom">
  <title>Montage Workflow</title>
  <author>   <name>Gaurang Mehta</name>
    <email>gmehta@isi.edu</email>   </author>
  <id>http://pegasus.isi.edu/workflows/montage/1</id>
  <link
href="http://pegasus.isi.edu/workflows/montage/montage-1-0.1.whip" rel="alternate"/>
  <updated>2009-07-30T23:19:03Z</updated>
  <summary>This workflow from the Montage (http://montage.ipac.caltech.edu) application is used to generate science…
…
  <category
scheme="http://org.whipplugin/data/description/datatype" term="http://pegasus.isi.edu/schema/DAX" label="The format of the workflow description"/>
  <category
scheme="http://org.whipplugin/data/description/entrypoint" term="data/montage.dax" label="The workflow description" />
  <category
scheme="http://pegasus.isi.edu/workflows/inputfile" term="data/input/2mass-atlas-990502s-j1420186.fits" size="2111040" label="An input file"/>
…
   <category
scheme="http://pegasus.isi.edu/workflows/VM" term="http://pegasus.isi.edu/workflows/montage/fc8-x86_64-montage.img" size="2684354560" arch="x86_64" os="Fedora Core 8" type="EC2 Image" label="The VM Image for Amazon EC2 containing Pegasus, Condor and Globus to run the workflow" />
  <category
scheme="http://pegasus.isi.edu/workflows/outputfile" term="data/output/mosaic.jpg" size="2478" label="An output file"/>
</entry>
```

**Figure 3: A snippet of the WHIP manifest.**

| Mosaic Size | Input Data | Intermediate Data | Output Data | Code | Workflow Specification | VM Size | Full Exec Dir | Total |
|---|---|---|---|---|---|---|---|---|
| 0.5 | 31.50MB | 251.7 MB | 56.3 MB | 49 MB | 0.081 MB | 759.8 MB | 1.2 MB | 1,149.581 MB |
| 1 | 94.5MB | 767 MB | 204 MB | 49 MB | 0.2607 MB | 759.8 MB | 3.2 MB | 1,877.7607 MB |
| 2 | 302.4MB | 2504 MB | 796 MB | 49 MB | 8769 MB | 759.8 MB | 11 MB | 1,3191.2 MB |
| 4 | 1224.3MB | 10300 MB | 3269 MB | 49 MB | 3.8 MB | 759.8 MB | 44 MB | 1,5649.9 MB |
| 6 | 2578.8 MB | 21938 MB | 7396 MB | 49 MB | 8 MB | 759.8 MB | 92 MB | 3,2821.6 MB |
| 8 | 4414.2 MB | 37951 MB | 13191 MB | 49 MB | 14 MB | 759.8 MB | 160 MB | 5,6539 MB |

Table 1: Data size of various Montage Workflow Artifacts.

## 5. EVALUATION

In order to evaluate the benefits of the proposed approach, we measured the amount of disk space needed to store provenance information in the traditional approach versus our pipeline provenance model. In the traditional approach all the information about the input, intermediate, and final data are stored as well as the workflow description and all the information associated with its execution.

Table 1 shows the disk space needed to store provenance information about Montage workflows of various sizes. As the size of the mosaics increases from 0.5 degree squares of the sky to 8 degree squares, so does the size of the input data and the size of the workflow to be executed. We distinguish between input data, intermediate data which are generated and consumed as part of the workflow execution, and the output data which correspond to the desired mosaic. The code represents the Montage code base, but does not include the workflow management code. The latter is included in the virtual image described below. The workflow specification is the size of the workflow as it is described in XML. The specification is in a form of a Directed Acyclic Graph, where the nodes of the graph represent the computations and their input and output data. The dependencies between the computations are also specified. The VM size corresponds to the size of the virtual machine image, which includes Fedora 8 with Java, Pegasus, Condor, Globus and some miscellaneous packages for Perl, C, C++, etc. This VM can be deployed on a cloud such as Amazon EC2 [23] or another virtual environment and be used as a host for the workflow computations. The full execution directory corresponds to all the files needed for the workflow engine to submit the workflow to the execution environment as well as all the logging information generated during the workflow execution.

In the traditional provenance model all the information above would be considered a part of the provenance record and would be stored. While this enables queries to be performed without the need for re-execution, this also adds significantly to the storage overhead.

In order to conduct a preliminary evaluation of our approach, we measured the data footprint of the traditional and pipeline provenance approach for the Montage application when managed by the Pegasus Workflow Management System. The results are shown in Figure 4. The X-axis shows the size of the Montage mosaic in degrees square. The Y-axis shows the total data footprint in megabytes (on a logarithmic scale). We plotted four different quantities: 1) the data footprint of the traditional provenance approach, which saves everything seen in Table 1 with the exception of the VM, 2) the same quantity as 1) but includes the Virtual Machine image, 3) the data footprint of the proposed pipeline-centric provenance model, and 4) the data footprint of the pipeline-centric model as implemented in the WHIP format, which compresses the elements of the bundle.

In the pipeline-centric provenance package we included:

- The Metadata which includes the creator of the bundle, date of creation, workflow format
- The workflow description
- The input data
- A reference to the VM image

We can see that in this case the pipeline-centric approach is on average 70% more efficient in terms of storage than the traditional approach. Additionally, when the pipeline-centric approach is implemented as a WHIP bundle, this improvement grows to almost 90% (although one could argue that the traditional provenance records can be compressed as well). If we include only references to the input data rather than the data themselves, as can be done for applications that access well maintained data archives, then the bundles would be even smaller.

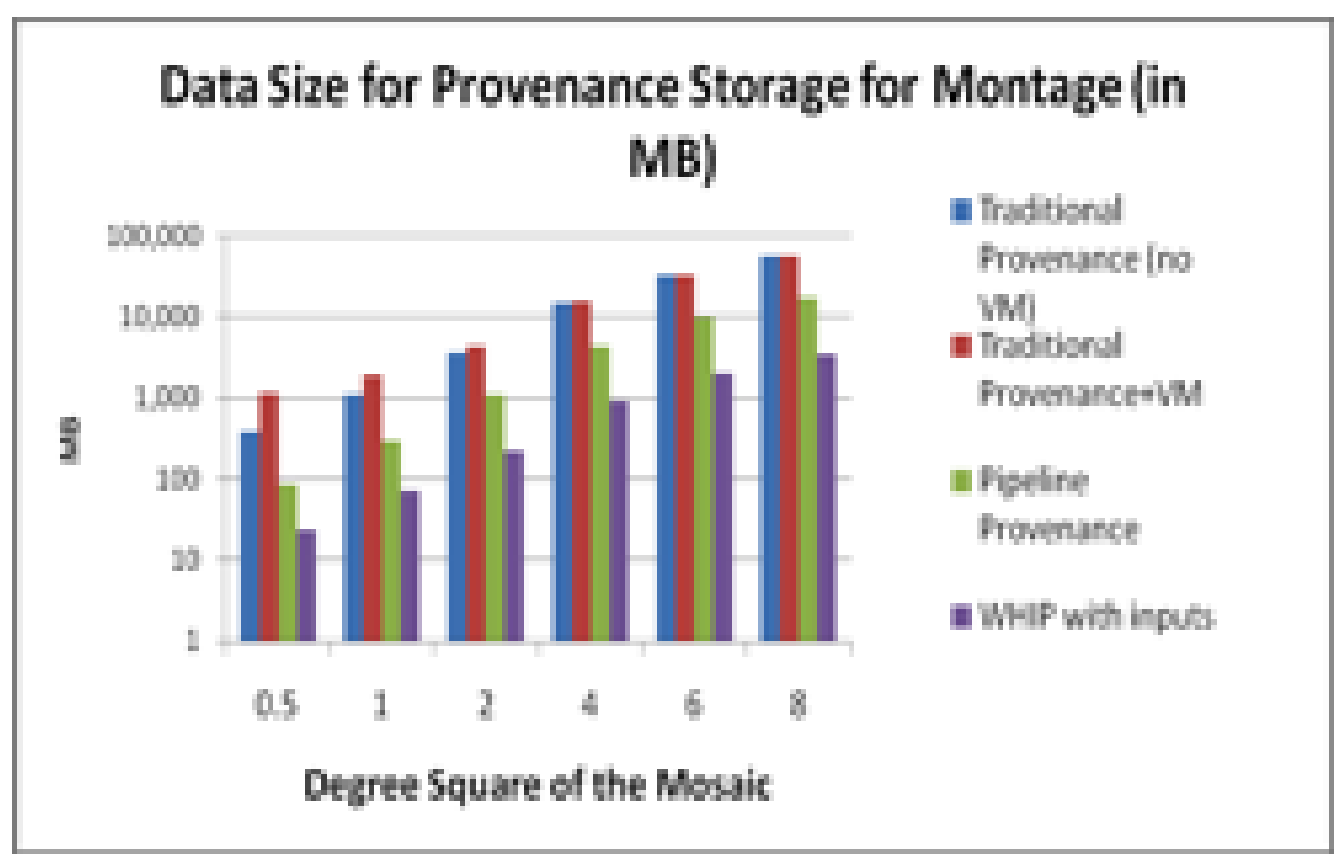

**Figure 4: Data Footprint of Different Approaches to Provenance.**

A drawback of our approach is that having only partial provenance records requires the workflow to be re-executed when a user wants to inspect or query the records. Thus we have the classic space versus time tradeoff. In our model, we assume that provenance data will not be frequently inspected or queried and thus re-execution will not be expensive. However, if some data are more popular than others, it may be beneficial to keep their full provenance records to be able to efficiently answer provenance queries.

In order to quantify the cost (in time) of workflow re-execution, we show the runtime of Montage on an Amazon EC2 extra large, 64-bit, high CPU instance with 7 GB of memory, 8 virtual cores with 2.5 EC2 Compute Units each, 1690 GB of instance storage, and high I/O performance. The cost of such an instance is $0.80 per instance hour (http://aws.amazon.com/ec2/instance-types/). Figure 4 shows the runtime of Montage on such an instance.

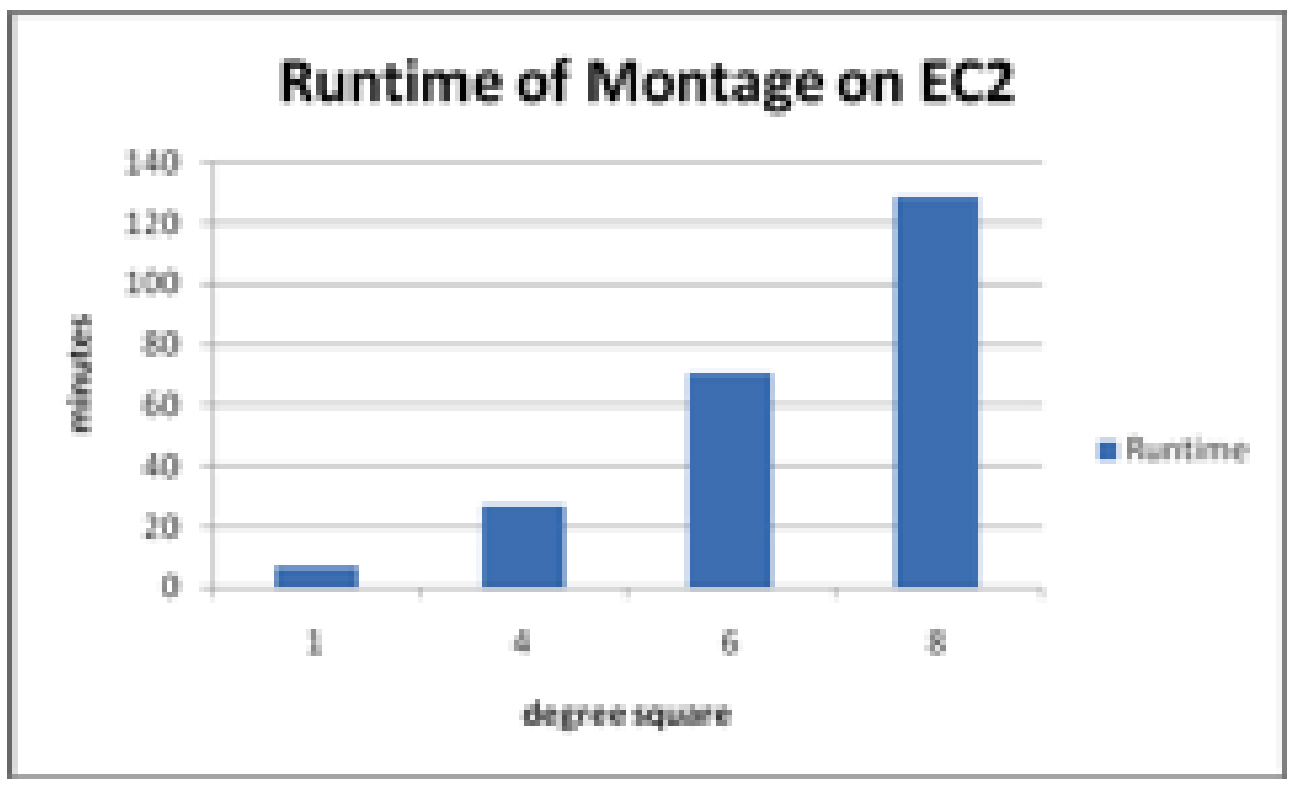

**Figure 4: Runtime of Montage on a Large Instance of EC2.**

For the largest size mosaic (8 degrees square) the runtime is just over 2 hours on the average, for a cost of $2.40. Generating 2 or 4 degree square mosaic takes less than 30 minutes and costs $0.80.

## 6. RELATED WORK

A number of systems and approaches have been developed to address provenance in e-Science applications. Bose and Frew [3] provide an extensive overview of provenance systems. Simmahn et al. [4] discuss various provenance systems for use in e-Science. Some systems, such as PASS [24], are execution-centric, focusing on gathering runtime information in the context of interactive applications. Other systems, such as Taverna [25], use a workflow to organize provenance information at runtime [26]. Finally, the database community has focused on the provenance of derived tuples. A good example of an extended database is Trio [27].

Unlike database systems, the applications we consider run on heterogeneous data (usually in the form of files) using complex codes. Execution-centric systems cater to more interactive applications whereas the applications we consider are well defined at the outset. The pipeline centric provenance model is closest to workflow-centric models. However, these models are not focused on re-execution as the mechanism to retrieve provenance. Instead, they use the workflow as a way to structure provenance information.

The closest work to our approach is the Virtual Data System (VDS) [28]. In this approach, provenance queries are, in some cases, answered by re-executing a workflow to retrieve intermediate results as suggested by the pipeline provenance model. Our work differs in that VDS uses a centralized database to store provenance data whereas our model focuses on collecting provenance information into an easily transportable package. Additionally, our work helps users understand when a provenance system based on re-execution is appropriate for their application. Lastly, unlike VDS, we focus on the storage advantages of this approach.

Other work has considered how to efficiently store provenance information. Chapman et al. describe a series of "provenance factorization" algorithms that find common subtrees in a provenance graph, which can be then collapsed to reduce the size of the provenance graph [29]. Heinis and Alonso describe an interval representation for provenance graphs that significantly reduces their size [30]. Groth et al. describe the use of references to reduce the size of provenance graphs [31]. The pipeline provenance model differs from these approaches in that it uses the notion of reproducibility to compress provenance information. However, unlike these methods, our approach may have significant query time impact because of the need for re-execution to retrieve intermediate data.

The concept of reproducibility has been discussed widely as a motivating factor for provenance [32]. There is a broad movement to encourage reproducible science (http://www.rrplanet.com). Our approach is not just to use provenance for reproducibility, but use the notion of reproducibility as the basis for provenance capture.

## 7. CONCLUSIONS

The pipeline-centric provenance model provides a packaging mechanism to capture the provenance of data produced by a class of applications common in e-Science. While this model can capture the data necessary to cover our example use-cases from Montage there is still work to be done to enable querying of these packages. We plan to add a query mechanism that transparently re-executes workflows to determine the provenance of intermediate data products. This query mechanism will support the retrieval of remotely stored data. We envision that the results of provenance queries will be returned as an Open Provenance Model graph [33] enabling interoperability between pipeline-centric provenance packages and other provenance systems. Once this query mechanism has been developed, we aim to measure the overhead of query by re-execution in comparison to standard query mechanisms that store all intermediate data. This will allow for the characterization of the trade-off between storage overhead and query time.

Both provenance and reproducibility are fundamental parts of the scientific process. For a certain class of scientific applications, the ability to reproduce a result can provide enough information to determine the provenance of output data. In this paper, we have described a pipeline-centric provenance model that captures all necessary information for provenance in a single package. Furthermore, we have shown using an initial set of experiments that significant storage reductions can be achieved using this model. Finally, we have described the type of applications that are suitable for this model. This work is the first step towards a greater understanding of the intersection of provenance and reproducibility in scientific workflow- based applications.

## 8. ACKNOWLEDGMENTS

The authors would like to acknowledge the support of the National Science Foundation under the SciFlow grant (CCF-0725332).